\documentclass[twocolumn, showpacs,aps,superscriptaddress,prl]{revtex4}

\usepackage{graphicx}
\usepackage{dcolumn}
\usepackage{bm}

\newcommand{\eq}[1]{(\ref{#1})}

\begin{document}


\title{Momentum injection in tokamak plasmas and transitions to reduced transport}

\author{F. I. Parra}
\email{f.parradiaz1@physics.ox.ac.uk} \affiliation{Rudolf Peierls
Centre for Theoretical Physics, University of Oxford, Oxford OX1
3NP, UK} \affiliation{Isaac Newton Institute for Mathematical
Sciences, Cambridge CB3 0EH, UK}
\author{M. Barnes}
\affiliation{Rudolf Peierls Centre for Theoretical Physics,
University of Oxford, Oxford OX1 3NP, UK}
\affiliation{Euratom/CCFE Fusion Association, Culham Science
Centre, Abingdon OX14 3DB, UK} \affiliation{Isaac Newton Institute
for Mathematical Sciences, Cambridge CB3 0EH, UK}
\author{E. G. Highcock}
\affiliation{Rudolf Peierls Centre for Theoretical Physics,
University of Oxford, Oxford OX1 3NP, UK} \affiliation{Isaac
Newton Institute for Mathematical Sciences, Cambridge CB3 0EH, UK}
\author{A. A. Schekochihin}
\affiliation{Rudolf Peierls Centre for Theoretical Physics,
University of Oxford, Oxford OX1 3NP, UK} \affiliation{Isaac
Newton Institute for Mathematical Sciences, Cambridge CB3 0EH, UK}
\author{S. C. Cowley}
\affiliation{Euratom/CCFE Fusion Association, Culham Science
Centre, Abingdon OX14 3DB, UK} \affiliation{Isaac Newton Institute
for Mathematical Sciences, Cambridge CB3 0EH, UK}

\date{\today}

\begin{abstract}
The effect of momentum injection on the temperature gradient in
tokamak plasmas is studied. A plausible scenario for transitions
to reduced transport regimes is proposed. The transition happens
when there is sufficient momentum input so that the velocity shear
can suppress or reduce the turbulence. However, it is possible to
drive too much velocity shear and rekindle the turbulent
transport. The optimal level of momentum injection is determined.
The reduction in transport is maximized in the regions of low or
zero magnetic shear.
\end{abstract}

\maketitle

\emph{Introduction.} In this Letter, we study the effect of
velocity shear on turbulent transport in tokamaks to answer two
questions: (a) what is the optimal momentum input that minimizes
radial energy transport? and (b) under what conditions do abrupt
transitions to reduced transport regimes occur? Experimentally,
tokamak plasmas can develop regions of reduced transport where the
temperature gradient is much higher than the typical value for the
same energy input \cite{connor04}, leading to more stored energy
and better performance at less cost. In large tokamaks, these
Internal Transport Barriers (ITBs) are found in regimes with low
magnetic shear and with a net momentum input by neutral beams
\cite{deVries09a, deVries09b}. In previous work, flow shear
\cite{dorland94, staebler94, barnes10, highcock10} and the
Shafranov shift \cite{beer97} have been proposed as causes for the
transition to reduced transport. Here we highlight the physical
influence of the velocity shear and momentum input. Employing
basic properties of the turbulent transport deduced from new
numerical results \cite{barnes10, highcock10} we show that there
is an optimal level of momentum input, and prove that abrupt
transitions to reduced transport are possible because the steady
state transport equations have several solutions, allowing for
bifurcations. We also obtain the conditions under which
transitions can occur.

\begin{figure}
\includegraphics[width = 6 cm, height = 4cm]{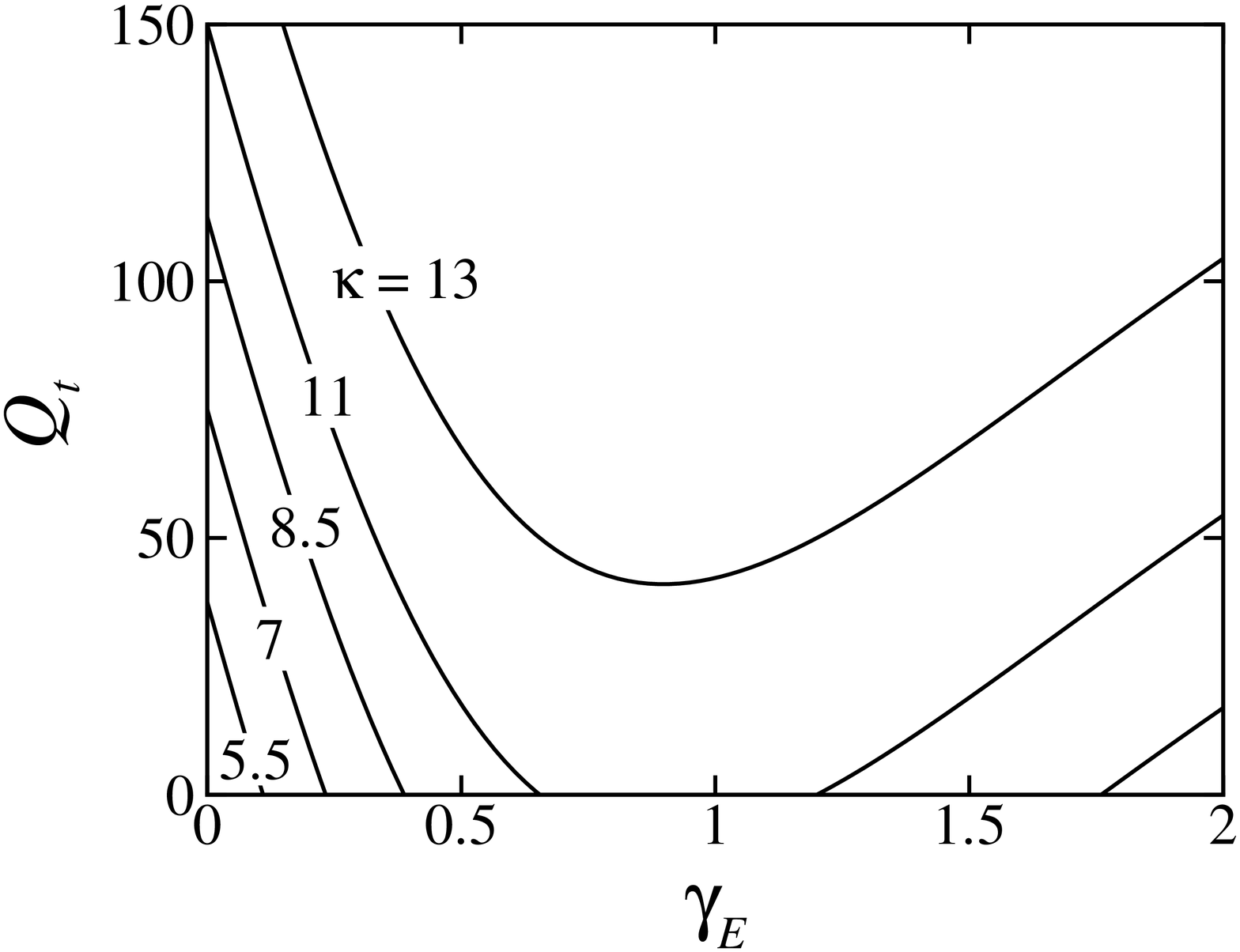}

\caption{\label{fig_Q} Schematic dependence of the turbulent
energy flux $Q_t$ on the velocity shear $\gamma_E$ and the
temperature gradient $\kappa$.}
\end{figure}

\emph{State of numerical evidence.} Refs.~\cite{barnes10,
highcock10} studied numerically the effect of flow shear on the
turbulent ion radial energy flux $Q_t$ with finite \cite{barnes10}
and zero \cite{highcock10} magnetic shear. In Fig.~\ref{fig_Q}, we
sketch the dependence of $Q_t$ on the dimensionless parameters
$\kappa = R/L_T$ and $\gamma_E = (B_P/B_T) (R^2/v_{ti}) |\partial
\omega/\partial r|$, where $L_T = |d(\ln T_i)/dr|^{-1}$ is the
scale of variation of the ion temperature, $\omega$ is the
rotation rate, $v_{ti}$ is the ion thermal speed, $r$ and $R$ are
the minor and major radius, and $B_T$ and $B_P$ are the toroidal
and poloidal magnetic field. The energy flux is normalized by the
gyroBohm value $Q_{gB} = (\rho_i/R)^2 p_i v_{ti}$, with $p_i$ and
$\rho_i$ the ion pressure and gyroradius. The curves in
Fig.~\ref{fig_Q} are generated by a simple analytic model chosen
to approximate the zero magnetic shear results of
\cite{highcock10}. For every $\kappa$, there is a minimum $Q_t$,
and for sufficiently small $\kappa$, this minimum is zero. For
large $\gamma_E$, the parallel velocity gradient drives an
instability that rekindles the turbulence \cite{catto73,
newton10}. The dependence of $Q_t$ on $\kappa$ and $\gamma_E$ is
qualitatively similar for finite magnetic shear \cite{barnes10},
but quantitatively there is a considerable difference: with zero
magnetic shear, the minima in $Q_t$ are much smaller for the same
$\kappa$, and the region of $\gamma_E$ for which the turbulence is
suppressed is wider.

The turbulent flux of toroidal angular momentum $\Pi_t$ was also
calculated in \cite{barnes10, highcock10}. It is also normalized
by the gyroBohm value $\Pi_{gB} = (\rho_i/R)^2 R p_i$. The
dependence of $\Pi_t$ on $\kappa$ and $\gamma_E$ has a remarkable
property: defining the normalized turbulent diffusivities as
$\chi_t = Q_t/\kappa$ and $\nu_t = \Pi_t/[(B_T/B_P) \gamma_E]$,
the turbulent Prandtl number $Pr_t = \nu_t/\chi_t$ was found to be
approximately independent of $\kappa$ and $\gamma_E$ and of order
unity \footnote{The simulations in \cite{barnes10, highcock10}
were for an up-down symmetric tokamak and near sonic flow. For
large up-down asymmetry \cite{camenen09b} or for very subsonic
flow \cite{parra10a}, $\Pi_t$ can be driven by temperature and
density gradients, and the Prandtl number defined here will not be
a constant.}.

\emph{Graphical analysis}. We analyze a plasma heated by neutral
beams. Consider a flux surface that contains the volume inside
which the energy and momentum are deposited. The ratio of the
injected momentum and energy fluxes is $\Pi_b/Q_b \sim C
v_{ti}/V_b$, where $V_b$ is the beam velocity, $C$ is a
geometrical constant dependent on the angle of the beams and
$\Pi_b$ and $Q_b$ are normalized by the gyroBohm values. Thus,
$\Pi_b/Q_b$ is a constant that only depends on the characteristics
of the beam. In experiments, $\Pi_b/ Q_b$ is usually of the order
of 0.1 \cite{mckee09}.

\begin{figure}
\includegraphics[width = 8.3 cm, height = 5cm]{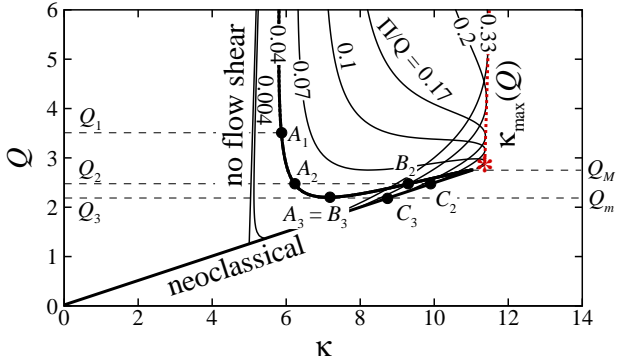}

\caption{\label{fig_Q_total} Energy flux $Q$ vs. temperature
gradient $\kappa$ for a constant ratio $\Pi/Q$ of momentum and
energy input.}
\end{figure}

To determine $\kappa$ and $\gamma_E$, we need to solve the
equations $Q = Q_t + Q_n = Q_b$ and $\Pi = \Pi_t + \Pi_n = \Pi_b$
\footnote{The effect of a turbulent inward pinch of momentum
\cite{peeters07} can be modelled by modifying the momentum input
$\Pi = \Pi_b + \mathrm{pinch}$. As a result, $\Pi/Q
> \Pi_b/Q_b$.}, where $Q_n = \chi_n \kappa$ and $\Pi_n = \nu_n
(B_T/B_P) \gamma_E$ are the collisional {\it neoclassical} energy
and momentum fluxes \cite{hinton85}. The dimensionless
diffusivities $\chi_n$ and $\nu_n$ are proportional to the ion-ion
collision frequency and depend on the magnetic field geometry.
Importantly, the neoclassical Prandtl number $Pr_n = \nu_n/\chi_n
\sim 0.1$ is smaller than the turbulent Prandtl number $Pr_t \sim
1$ \cite{hinton85}.

\begin{figure}
\includegraphics[width = 7.5 cm, height = 11cm]{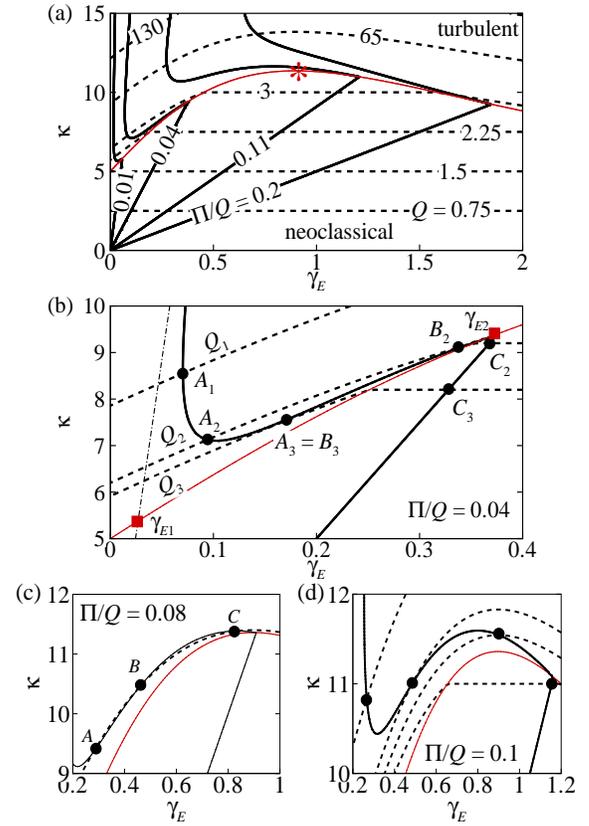}

\caption{\label{fig_contour} (a) Curves of constant $Q$ (dashed
lines) and constant $\Pi/Q$ (solid lines). The thin red line is
the critical temperature gradient $\kappa_c$ below which there is
no turbulence. (b) Sketch of the intersection between a curve of
given $\Pi/Q$ and curves of constant $Q$ with $Q_1 > Q_2> Q_3$.
The black dash-dot line is Eq.~\eq{PiQ_t}. (c), (d) Similar
sketches for higher $\Pi/Q$.}
\end{figure}

To describe the solutions to $Q = Q_b$ and $\Pi = \Pi_b$, we plot
the curves of constant $\Pi/Q$ in a $(\kappa, Q)$ graph, as given
in Fig.~\ref{fig_Q_total} \footnote{Fig.~4(b) of \cite{highcock10}
is a numerical reconstruction of Fig.~\ref{fig_Q_total}.}. The
beam characteristics determine the curve of constant $\Pi/Q$.
Then, given $Q$, the temperature gradient $\kappa$ is easy to read
off the graph.

To understand Fig.~\ref{fig_Q_total}, it is convenient to consider
the $(\gamma_E, \kappa)$ parameter space and search for points of
intersection of curves of constant $Q$ and $\Pi/Q$. From the
general shape of the constant $\kappa$ curves in Fig.~\ref{fig_Q},
we infer the contours of constant $Q$ in the $(\gamma_E, \kappa)$
plane, shown in Fig.~\ref{fig_contour}(a). The transport is purely
neoclassical for $\kappa$ below the critical value
$\kappa_c(\gamma_E)$ (this corresponds to the points in
Fig.~\ref{fig_Q} where $Q_t$ vanishes; it is \emph{not} the linear
stability boundary \footnote{For large $\gamma_E$, the plasma is
linearly stable \cite{highcock10}. Subcritical turbulence exists
because small perturbations grow transiently due to temperature
and parallel velocity gradients. For turbulence to exist, $\kappa$
must still be larger than some $\kappa_c$.}). At $\kappa <
\kappa_c$, the constant $Q$ curves are horizontal because $Q_n$ is
independent of $\gamma_E$. At $\kappa> \kappa_c$, since
neoclassical transport is usually much smaller than turbulent
transport, the constant $Q$ curves are approximately the constant
$Q_t$ curves. We stress that $\kappa_c (\gamma_E)$ is the curve of
$Q_t = 0$, but it is \emph{not} a curve of constant $Q = Q_t +
Q_n$. In Fig.~\ref{fig_contour}(a), we have exaggerated the
difference.

The curves of constant $\Pi/Q$ are also shown in
Fig.~\ref{fig_contour}(a). For $\kappa < \kappa_c$, the transport
is neoclassical, and for $\kappa \gg \kappa_c$, turbulence
dominates. Therefore
\begin{eqnarray}
\kappa = (\Pi/Q)^{-1} Pr_n (B_T/B_P) \gamma_E \quad \mathrm{for}
\; \kappa < \kappa_c, \label{PiQ_n} \\ \kappa = (\Pi/Q)^{-1} Pr_t
(B_T/B_P) \gamma_E  \quad \mathrm{for} \; \kappa \gg \kappa_c.
\label{PiQ_t}
\end{eqnarray}
In both regimes, the curves of constant $\Pi/Q$ are straight lines
passing through the origin. Since $Pr_t > Pr_n$, these lines are
steeper in the turbulent than in the neoclassical regime. To
transit from the former to the latter, the curves of constant
$\Pi/Q$ must approximately follow the curve $\kappa_c (\gamma_E)$
because neoclassical and turbulent transport are comparable in its
vicinity. In Fig.~\ref{fig_contour}(a), the transition from
Eq.~\eq{PiQ_n} to Eq.~\eq{PiQ_t} is shown in detail by
exaggerating the difference between the constant $\Pi/Q$ curve and
$\kappa_c(\gamma_E)$. This piece of the curve of constant $\Pi/Q$
is crucial for bifurcations.

Fig.~\ref{fig_Q_total} was produced using
Fig.~\ref{fig_contour}(a). The intersections of constant $Q$ and
constant $\Pi/Q$ curves for $\kappa \gg \kappa_c$ correspond to
the high $Q$ section of the curves in Fig.~\ref{fig_Q_total}, and
the intersections for $\kappa < \kappa_c$ form the neoclassical
straight line. The region in between, where for each value of $Q$
and $\Pi$ we can find several values of $\kappa$, is examined in
the $(\gamma_E, \kappa)$ space below in the section on
bifurcations.

\emph{Optimal momentum injection}. In Fig.~\ref{fig_Q_total}, it
is clear that to maximize $\kappa$, we need to operate on the red
dashed line $\kappa_\mathrm{max} (Q)$ that corresponds to the the
maxima in $\kappa$ at constant $Q$ in Fig.~\ref{fig_contour}(a).
However, once there, any increase in $\kappa$ achieved by
increasing the energy input $Q$ is small because turbulent
transport is very stiff. Therefore, the optimal operation is at
the maximum critical temperature gradient, $\kappa_{c,
\mathrm{max}} = \kappa_c (\gamma_{E, \mathrm{max}})$, given in
Fig.~\ref{fig_Q_total} and Fig.~\ref{fig_contour}(a) as a red
star. As a result, the optimal temperature gradient is $\kappa_{c,
\mathrm{max}}$, the optimal momentum input is $\Pi/Q = (B_T/B_P)
Pr_n (\gamma_{E, \mathrm{max}}/\kappa_{c, \mathrm{max}})$, and the
optimal energy flux is $Q = \chi_n \kappa_{c, \mathrm{max}}$.

\emph{Conditions for bifurcations}. Transitions can happen only
when there are several values of $\kappa$ and $\gamma_E$ for given
values of $Q$ and $\Pi$. The curves of constant $Q$ and constant
$\Pi/Q$ can intersect in multiple points, as exemplified by the
thicker line in Fig.~\ref{fig_Q_total}. Fig.~\ref{fig_contour}(b)
is a sketch of the curves in the $(\gamma_E, \kappa)$ plane that
correspond to this case. For $Q = Q_1$, turbulence dominates and
there is only one solution, $A_1$. If we decrease the energy input
to $Q_2$, there are three solutions $A_2$, $B_2$ and $C_2$, where
$C_2$ is neoclassical. A jump from $A_2$ to $C_2$ reduces the
transport and increases the temperature gradient. If we continue
decreasing $Q$ to $Q_3$, the constant $Q$ and $\Pi/Q$ curves
become tangent and there are two solutions. For $Q < Q_3$, there
is only one solution, which is neoclassical.

\begin{figure}
\includegraphics[width = 7.5 cm, height = 11cm]{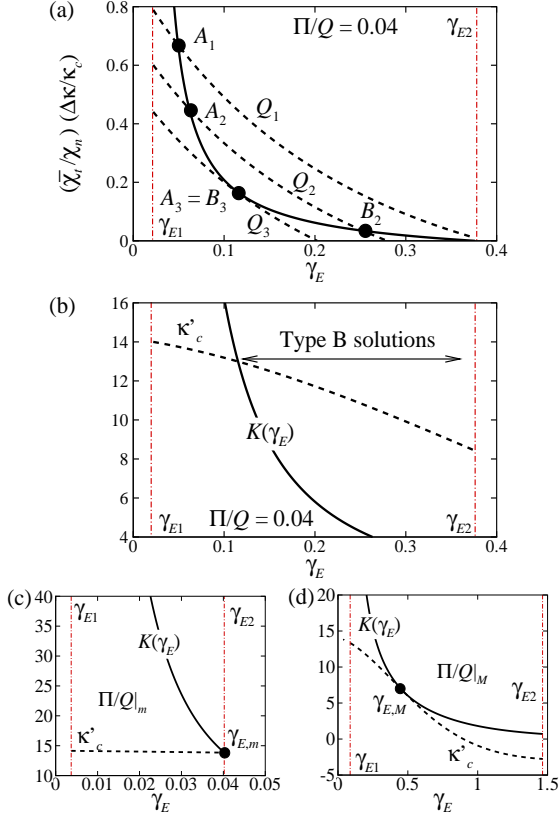}

\caption{\label{fig_inter_2} (a) Asymptotic approximation to the
curves of constant $Q$ (dashed lines) and constant $\Pi/Q$ (solid
lines) in the transition region between neoclassical and turbulent
regimes. (b) Graphical representation of condition \eq{local_cond}
for type $B$ solutions to exist. (c), (d) The cases in which type
$B$ solutions no longer exist for small and large $\Pi/Q$.}
\end{figure}

Increasing $\Pi/Q$ above the value in Fig.~\ref{fig_contour}(b)
gives the curves in Fig.~\ref{fig_contour}(c). The solution $C$
with the largest $\kappa$ is not purely neoclassical because the
large momentum input causes a large parallel velocity gradient,
which drives turbulence \cite{barnes10, highcock10, catto73,
newton10}. At even larger $\Pi/Q$, the situation is as in
Fig.~\ref{fig_contour}(d), where there is only one solution for
each $Q$ and bifurcations are not possible. It is easy to see how
these cases are reflected in the large $\Pi/Q$ curves of
Fig.~\ref{fig_Q_total}. We now discuss the conditions for several
solutions to exist.

Compare Figs.~\ref{fig_contour}(b) and \ref{fig_contour}(d). The
existence of several solutions is determined by the slope of the
piece of the curve of constant $\Pi/Q$ that transits between the
neoclassical and turbulent regimes. We study that region to prove
that to have several solutions and hence transitions, $\Pi/Q$ and
$Q$ must be within a domain determined by the shape of the curve
$\kappa_c (\gamma_E)$. Near this critical curve, $Q_t \simeq
\overline{\chi}_t (\gamma_E) [ \kappa - \kappa_c (\gamma_E) ]$,
where $\overline{\chi}_t = \kappa_c (\partial \chi_t/\partial
\kappa)|_{\kappa = \kappa_c}$ and $\Delta \kappa = \kappa -
\kappa_c \ll \kappa_c$. We also assume that $Pr_t$ remains
approximately constant even for $\kappa \simeq \kappa_c$. Then,
expanding in $\chi_n/\overline{\chi}_t \ll 1$ and ordering $Pr_t
\sim Pr_n \sim 1$, we find the curves of constant $Q$ and $\Pi/Q$
to be
\begin{eqnarray}
 & \Delta \kappa_Q (\gamma_E) = \frac{\mbox{$Q$}}{\mbox{$\overline{\chi}_t$}} -
\frac{\mbox{$\chi_n \kappa_c$}}{\mbox{$\overline{\chi}_t$}}, & \\
 & \Delta \kappa_{\Pi/Q} (\gamma_E) = \frac{\mbox{$\chi_n
\kappa_c$}}{\mbox{$\overline{\chi}_t$}} \frac{\mbox{$\Pi/Q - Pr_n
(B_T/B_P) (\gamma_E/\kappa_c)$}}{\mbox{$Pr_t (B_T/B_P)
(\gamma_E/\kappa_c) - \Pi/Q$}}. &
\end{eqnarray}
We plot these approximate expressions in
Fig.~\ref{fig_inter_2}(a). The expression for $\Delta
\kappa_{\Pi/Q} (\gamma_E)$ is only valid for the transition region
between the turbulent and neoclassical regimes, i.e., for
$\gamma_{E1} < \gamma_E < \gamma_{E2}$, where $\gamma_{E1}$ and
$\gamma_{E2}$ are the intersections between the curve
$\kappa_c(\gamma_E)$ and the lines \eq{PiQ_n} and \eq{PiQ_t},
i.e., $Pr_t (B_T/B_P) [\gamma_{E1}/\kappa_c(\gamma_{E1})] = \Pi/Q$
and $Pr_n (B_T/B_P) [\gamma_{E2}/\kappa_c(\gamma_{E2})] = \Pi/Q$.
These points of intersection are marked as red squares in
Fig.~\ref{fig_contour}(b), and as red dash-dot lines in
Fig.~\ref{fig_inter_2}(a).

The solutions to $Q = Q_b$ and $\Pi/Q = \Pi_b/Q_b$ are given by
$\Delta \kappa_Q (\gamma_E) = \Delta \kappa_{\Pi/Q} (\gamma_E)$.
In Fig.~\ref{fig_inter_2}(a), we recast Fig.~\ref{fig_contour}(b)
in terms of $\Delta \kappa$. To have several solutions we need
type B solutions that we define as intersections where $\Delta
\kappa_Q^\prime (\gamma_{E, B}) < \Delta \kappa_{\Pi/Q}^\prime
(\gamma_{E, B})$. Here $^\prime$ denotes differentiation with
respect to $\gamma_E$. This condition gives
\begin{equation}
\kappa_c^\prime(\gamma_{E,B}) > \frac{1}{Pr_t} \frac{\Pi}{Q}
\frac{B_P}{B_T} \left [ \frac{\kappa_c
(\gamma_{E,B})}{\gamma_{E,B}} \right ]^2 \equiv K (\gamma_{E,B}).
\label{local_cond}
\end{equation}
In Fig.~\ref{fig_inter_2}(b), the dashed line is
$\kappa_c^\prime$, and the solid line is $K (\gamma_E)$. Condition
\eq{local_cond} is never satisfied near $\gamma_{E1}$ because
$\Delta \kappa_{\Pi/Q}^\prime \rightarrow - \infty$ there.

\begin{figure}
\includegraphics[width = 7 cm, height = 3.8cm]{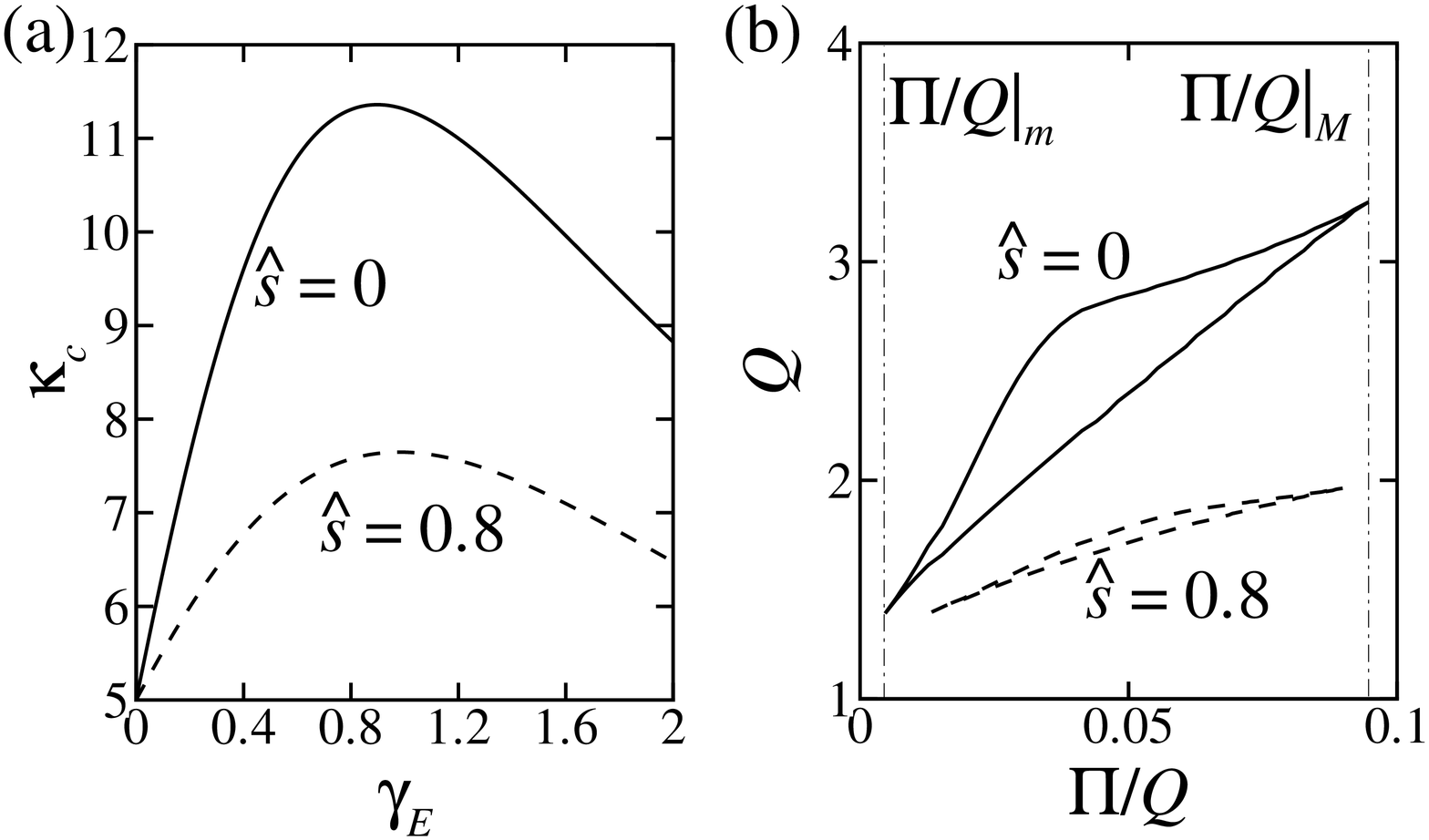}

\caption{\label{fig_limits} (a) Critical temperature gradient
$\kappa_c$ for zero (solid line) and finite (dashed line) magnetic
shear $\hat{s}$ based on \cite{highcock10} and \cite{barnes10}.
(b) Region in the $(\Pi/Q, Q)$ space where abrupt transitions can
happen.}
\end{figure}

Condition \eq{local_cond} defines an interval
\begin{equation}
\left. \frac{\Pi}{Q} \right |_m < \frac{\Pi}{Q} < \left.
\frac{\Pi}{Q} \right |_M
\end{equation}
outside of which there is only one solution for each $Q$.
Inequality \eq{local_cond} is plotted for $\Pi/Q = \Pi/Q|_m$ and
$\Pi/Q = \Pi/Q|_M$ in Figs.~\ref{fig_inter_2}(c) and
\ref{fig_inter_2}(d), respectively. At both $\gamma_{E,m}$ and
$\gamma_{E,M}$, $\kappa_c^\prime = K$. In addition, $\gamma_{E,m}
= \gamma_{E2}$, whereas at $\gamma_{E,M}$,
$\kappa_c^{\prime\prime} = K^\prime$. Then, $\gamma_{E,m}$ and
$\gamma_{E,M}$ are given by
\begin{eqnarray}
 & \kappa_c^\prime (\gamma_{E, m}) = \frac{\mbox{$Pr_n$}}{\mbox{$Pr_t$}}
\frac{\mbox{$\kappa_c(\gamma_{E,m})$}}{\mbox{$\gamma_{E,m}$}}, & \\
 & \kappa_c^{\prime \prime} (\gamma_{E,M}) = \frac{\mbox{$2 \kappa_c^\prime
(\gamma_{E,M})$}}{\mbox{$\gamma_{E,M}$}} \left [
\frac{\mbox{$\gamma_{E,M} \kappa_c^\prime
(\gamma_{E,M})$}}{\mbox{$\kappa_c (\gamma_{E,M})$}} - 1 \right ].
&
\end{eqnarray}
Once $\gamma_{E,m}$ and $\gamma_{E,M}$ are known, $\Pi/Q|_m$ and
$\Pi/Q |_M$ can be obtained from $\kappa_c^\prime (\gamma_E) =
K(\gamma_E)$, which leads to $\Pi/Q = Pr_t (B_T/B_P)
\kappa_c^\prime (\gamma_E) [ \gamma_E/\kappa_c (\gamma_E) ]^2$.
The lower limit $\Pi/Q|_m$ appears because $K(\gamma_E)$ is
bounded by its values at $\gamma_{E2}$,
$(Pr_n^2/Pr_t)(B_T/B_P)(\Pi/Q)^{-1}$, and at $\gamma_{E1}$, $Pr_t
(B_T/B_P)(\Pi/Q)^{-1}$, tending to infinity for $\Pi/Q \rightarrow
0$. The upper limit $\Pi/Q|_M$ arises because increasing $\Pi/Q$
shifts the interval $\gamma_{E1} < \gamma_E < \gamma_{E2}$ towards
higher values of $\gamma_E$ and eventually $\kappa_c^\prime$
becomes negative.

For every $\Pi/Q$, there is also an interval in $Q$,
\begin{equation}
Q_m < Q < Q_M,
\end{equation}
for which multiple solutions exist. We show the $Q$ limits for
$\Pi/Q = 0.04$ in Fig.~\ref{fig_Q_total}. The lower limit $Q_m$ is
$Q_3$ and the upper limit $Q_M$ is the meeting point with
neoclassical transport.

In Fig.~\ref{fig_limits}(b), we give the domain in the $(\Pi/Q,
Q)$ plane where for each $Q$ and $\Pi/Q$ there are several
solutions for $\kappa$ and $\gamma_E$. We do so for zero
\cite{highcock10} and finite \cite{barnes10} magnetic shear, whose
$\kappa_c (\gamma_E)$ curves are given in
Fig.~\ref{fig_limits}(a). The derivative $\kappa_c^\prime$ is
clearly smaller for finite magnetic shear because the velocity
shear is less efficient in quenching the turbulence. As a result,
as we expect from \eq{local_cond}, the region with several
solutions is smaller. Thus, transitions are more probable at zero
or small magnetic shear.

\emph{Conclusions.} In tokamaks, the turbulent transport of energy
and momentum satisfies two properties: (i) given the energy flux
$Q$, the temperature gradient $\kappa$ has a maximum possible
value achieved at a finite flow shear $\gamma_E$; and (ii) the
turbulent Prandtl number $Pr_t$ is approximately constant. Using
these properties we have found the optimal level of momentum
input. In addition, employing the fact that the neoclassical
Prandtl number $Pr_n$ is smaller than $Pr_t$, we have shown that
transitions to reduced transport can occur in an interval
$\Pi/Q|_m < \Pi/Q < \Pi/Q|_M$. Below $\Pi/Q|_m$ the flow shear is
not sufficient to suppress the turbulence, and above $\Pi/Q|_M$
the large parallel velocity gradient drives strong turbulence. For
each $\Pi/Q$, since bifurcations occur only when neoclassical and
turbulent transport are comparable and the effective Prandtl
number is not constant, $Q$ must be in the interval $Q_m < Q <
Q_M$. Below $Q_m$, the transport is neoclassical or close to
neoclassical, and above $Q_M$, it is mainly turbulent. The region
in $(\Pi/Q, Q)$ space where transitions occur is wider for small
magnetic shear, which is consistent with both numerical
\cite{barnes10, highcock10} and experimental \cite{deVries09a}
indications.

The authors are grateful to P. de Vries and G. Hammett for many
helpful discussions. This work was supported in part by EPSRC,
STFC and the Leverhulme Trust Network for Magnetized Plasma
Turbulence.

\end{document}